\documentclass[twocolumn,amsmath,amssymb,aps,prb,floatfix,longbibliography]{revtex4-1}

\usepackage{graphicx}% Include figure files
\usepackage{dcolumn}% Align table columns on decimal point
\usepackage{bm}% bold math

\usepackage{chemformula} % Formula subscripts using \ch{}
\usepackage[T1]{fontenc}       % Use modern font encodings
\usepackage{flexisym}		% For text \prime symbol
\usepackage{placeins}
\newcommand\T{\rule{0pt}{3.0ex}}
\newcommand\B{\rule[-1.2ex]{0pt}{0pt}}
\newcommand{\ra}[1]{\renewcommand{\arraystretch}{#1}}

\usepackage{booktabs}

\usepackage[normalem]{ulem}
\usepackage{pifont}
\newcommand{\mos}{\ch{MoS2}}
\newcommand{\ws}{\ch{WS2}}
\newcommand{\mose}{\ch{MoSe2}}
\newcommand{\wse}{\ch{WSe2}}

\newcommand{\mse}{\ch{MSe2}}
\newcommand{\ms}{\ch{MS2}}
\newcommand{\wx}{\ch{WX2}}
\newcommand{\mox}{\ch{MoX2}}
\newcommand{\Tp}{{T$^\prime$}}
\newcommand{\Tpp}{{T$^{\prime \prime}$}}

\begin{document}

\title{Composition Dependence of the Charge Driven Phase Transition in Group-VI Transition Metal Dichalcogenides}

\author{Urvesh Patil}
\author{Nuala M.~Caffrey}
\email{nuala.caffrey@tcd.ie}
\affiliation{School of Physics and CRANN$,$ Trinity College$,$ Dublin 2$,$ Ireland}

\date{\today}

\begin{abstract}
	
	Materials exhibiting multiple stable phases can be used as functional components in electronic and optical applications if the phase transition is controllable. Group-VI transition metal dichalcogenides (TMDs, MX$_2$, where M=Mo, W and X = S, Se) are known to undergo charge induced transitions from semi-conducting H phases to metallic T phases. This occurs, for example, when bulk TMDs are exfoliated with the aid of alkali ion intercalants. However, it is difficult to experimentally decouple the effect of composition-dependent phase transition barriers from indirect effects related to the exfoliation process. Here, using first-principles calculations, we study the energetics of transition between the different structural polytypes of four group-VI TMDs upon lithium adsorption. We find that both the activation barrier from the H phase to the metallic phase in charged monolayers and the reverse barrier in neutral monolayers are required to explain experimental results. We show that the high proportion of metallic phase found in WS$_2$ monolayers after alkali treatment can be explained by high transition barriers to revert back to the H phase once in a neutral state. The calculated barriers however cannot explain the low proportion of metallic phase found in MoS$_2$ monolayers in some experiments and so non-electronic effects must also play a role. 
	
\end{abstract}

\maketitle

\section{Introduction}
Transition metal dichalcogenides (TMDs), comprised of layered sheets of transition metal atoms sandwiched between two layers of chalcogen atoms (MX$_2$), are chemically versatile, exhibiting a broad range of electronic properties from insulating (ZrS$_2$) to superconducting (NbSe$_2$) \cite{manzeli20172d}. As well as by changing the composition, the conductivity of a TMD can be modified by inducing a structural transition between different polymorphs. Several such polymorphs exist, distinguished by the metal coordination of the chalcogen atoms. 
In the semi-conducting H phase, the chalogen atoms are AA-stacked so that the metal atoms occupy alternate trigonal-prismatic voids.  The metallic T phase, on the other hand, has a tetragonal symmetry, with the metal atoms occupying octahedral voids between AB-stacked chalcogen atoms. This particular phase can transform to a semi-metallic distorted octahedral phase, designated here \Tp.
{Recently, a further mixed phase, designated \Tpp, which can be viewed as a series of alternating H and \Tp\ phases, was theoretically predicted to be lower in energy than the \Tp\ phase for \mos\ } \cite{ma2016predicting}.

The ability to induce a transition in \mos\ from its ground state H phase to an octahedral phase via alkali metal intercalation has been known since the 1980s~\cite{py1983structural}. This phase transition was attributed to a transfer of charge from the intercalated atom to the TMD, and more specifically to the \textit{d}~states of the transition metal atom~\cite{Kertesz1984,chia2018layered}. Indeed, the entire family of group-VI TMDs, where M = Mo, W and X = S, Se and Te, can be manipulated to undergo phase transitions close to ambient conditions \cite{voiry2015phase}. This has been achieved using a variety of methods, including alkali metal adsorption \cite{gamble1974ionicity, py1983structural,wang2014atomic,lei2018direct}, the introduction of impurities or vacancies \cite{raffone2016mos2,kretschmer2017structural,pizzochero2017point}, electron or laser irradiation~\cite{kretschmer2017structural, lin2014atomic, cho2015phase, kang2014plasmonic, fan2015fast, guo2015probing} and electrostatic gating \cite{radisavljevic2013mobility,wang2017structural}.

The electronic properties of a TMD are evidently strongly dependent on its structural phase. In \mos\ the conductivity of the T phase was found to be up to $10^7$ times higher than that of the semiconducting H phase \cite{acerce2015metallic}. The ability to reversibly and reliably switch between these two phases would result in applications as monolayer-thick field effect transistors \cite{nourbakhsh2016mos2}, gas sensors~\cite{friedman2016dynamics} and catalysts~\cite{luxa2018cation}.
Such applications require precise control over the phase transition process so that a complete phase change to \Tp\ phase can be achieved and maintained.

Liquid phase exfoliation, with the aid of alkali metal intercalation, is a common and effective way of isolating TMD monolayers from the bulk on an industrial scale. In such experiments, the alkali metal atoms are first intercalated into the bulk TMD using an organolithium compound. The intercalated TMD is then solvated in a polar solvent. This results in the discharging of the monolayers, the deintercalation of the metal ion and the exfoliation of individual monolayers from the bulk materials. These layers are subsequently measured to have both H and T phases present. 

Of the group-VI TMDs, the phase transition in \mos\ has been investigated in detail. Yet, compared to other group-VI TMDs, the phase transition efficiency in \mos\ can be relatively low: a comprehensive side-by-side comparison of the experimental transition efficiency in four group-VI TMDs -- \mos, \mose, \ws\ and \wse\ -- found that \ws\ exhibited the largest increase in the proportion of the 1T phase compared to the starting 2H phase (i.e., the 1T/2H ratio), followed by \mose\ and \wse, and eventually \mos. \cite{ambrosi20152h} While the exact ratios were subsequently found to depend on the nature of the organolithium intercalant group used in the experiment, the general trend remained: \wx\ compounds display a higher proportion of T phase compared to \mox, for both X = S and Se~\cite{rohaizad20171t, tan2016aromatic}.

In these alkali metal induced exfoliation experiments, the measured phase transition efficiency will depend not only on the intrinsic composition-dependent free-energy barrier between the different phases, but also on the composition-dependent exfoliation efficiency of the chosen organolithium intercalant. Experimentally, it is difficult to decouple these two contributions. {As ion-assisted liquid phase exfoliation is the most suitable method of producing sufficient T phase TMDs on an industrial scale, a complete understanding the mechanism is essential to maximize the amount of metallic phase produced and to prevent a transition back to the H phase.} 

Previous computational investigations of the phase transition barrier have concentrated on \mos, looking primarily at the threshold charge density required to induce the phase transition. However, a disconcertingly wide range of values have been reported, including 0.35~e per formula unit (f.u.) \cite{li2016structural},  0.55~e per f.u.~\cite{ma2016predicting}, 0.78~e per f.u.~\cite{sun2016origin} and almost 2~e per f.u.~\cite{gao2015charge}. {The origin of this discrepancy is discussed in detail in the Appendix.} 

Here, using first-principle calculations, we determine the transition barriers between all possible polytypes of both the pristine and Li-adsorbed group-IV TMDs, with the aim of determining whether the composition dependence of experimental transition efficiency in four group-VI TMDs can be explained using the intrinsic barriers to the phase transitions alone.

\section{Computational Methods}

Density functional theory (DFT) calculations are performed using the {\sc vasp-5.4} code~\cite{vasp3,vasp4,vaspPAW2}. The optB86b-vdW exchange-correlation functional was used to account for long range dispersion interactions.\cite{vaspvdw1,vaspvdw5} This functional was previously shown to give accurate lattice parameters and energies for layered materials \cite{bjorkman2014testing,urveshmolecules} 
All calculations are performed with a cutoff energy of 500~eV for the plane wave basis set. A $\Gamma$ centered K-point grid of 11$\times$7$\times$1 is used to calculate the total energy of the H and \Tp\ phases, while a 11$\times$5$\times$1 grid is used for the \Tpp\ mixed phase.

The structures are relaxed until the force on each atom is less than 0.01~eV/\AA. The unit cell length in the direction normal to the plane is fixed at 25~\AA\ for  H and \Tp\ phase calculations, and 26~\AA\ for \Tpp\ mixed phase calculations and all  transition state calculations. This corresponds to a minimum vacuum of 18~\AA\ between repeating monolayers and the dipole corrections are applied.

Transition states are determined using the climbing image nudged elastic band (CI-NEB) method \cite{henkelman2000climbing,henkelman2000improved} with a spring constant of 5. All atoms in the transition state calculations are optimized so that the force on each atom is less than 0.05~eV/\AA.

\section{Results}

\subsection{Neutral Monolayers}

\begin{figure}
	\centering
	\includegraphics[width=\columnwidth]{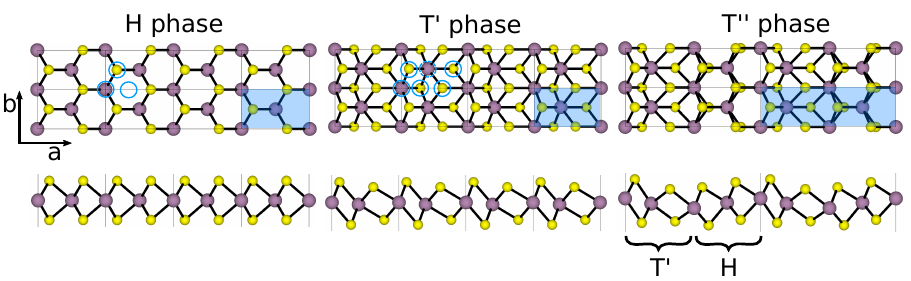}
	\caption{\label{fig:structures}  Top and side view of the H, \Tp\ and \Tpp\ polytypes. The orthorhombic cells of all three phases are shown as blue shaded regions. Note that the \Tpp\ mixed phase can be viewed as alternating \Tp\ and H phases. The blue circles indicate the possible candidate sites for lithium adsorption.}
\end{figure}

The crystal structures of the H, \Tp\ and \Tpp\ mixed phases are shown in Fig.~\ref{fig:structures}. The structural parameters determined for all neutral monolayers (without adsorbed Li) are given in Table~\ref{tab:structures}. The lattice constants are dictated by the chalcogen atom, with \mos\ and \ws\ having very similar values in all three phases. Likewise, \mose\ and \wse\ have almost identical lattice constants. These values are in good agreement with the available experimental data \cite{gordon2002structures,heising1999structure,wildervanck1964preparation,schonfeld1983anisotropic,bronsema1986structure,schutte1987crystal,dickinson1923crystal,bell1957preparation, podberezskaya2001crystal,samadi2018group} and with previous calculations\cite{kumar2012electronic,gusakova2017electronic,liu2013three,garcia2017tuning} in the literature. For the \Tp\ phase, the calculated $a$ lattice constant of 5.57~\AA\ for \mos\ is consistent with the experimentally observed length of 5.6~\AA.
\begin{table}\centering
	\ra{1.2}
	\setlength{\tabcolsep}{6pt} % Default value: 6pt
	\begin{tabular}{@{\extracolsep{3pt}}lcccccc@{}}
		\hline \hline 
		\T \B   & \multicolumn{2}{c}{H} &   \multicolumn{2}{c}{\Tp}  &   \multicolumn{2}{c}{\Tpp}  \\
		\cline{2-3} \cline{4-5} \cline{6-7}
		\T \B  & a  & b  & a & b  & a  & b \\
		\cline{2-3} \cline{4-5} \cline{6-7}
		\mos\ \T \B	&5.48&3.16&5.57&3.22&11.23&3.16\\
		\mose\ \T \B	&5.71&3.30&5.79&3.34&11.69&3.27\\
		\ws\ \T \B	&5.48&3.17&5.59&3.23&11.24&3.18\\
		\wse\ \T \B	&5.71&3.30&5.81&3.35&11.69&3.28\\
		\hline
	\end{tabular}
	\caption{\label{tab:structures}\,The lattice parameters a, b (in \AA) of the H, the \Tp, and the  \Tpp\ mixed phase of \mos, \mose, \ws\ and \wse.}
\end{table}

\begin{figure}
\includegraphics[width=\columnwidth]{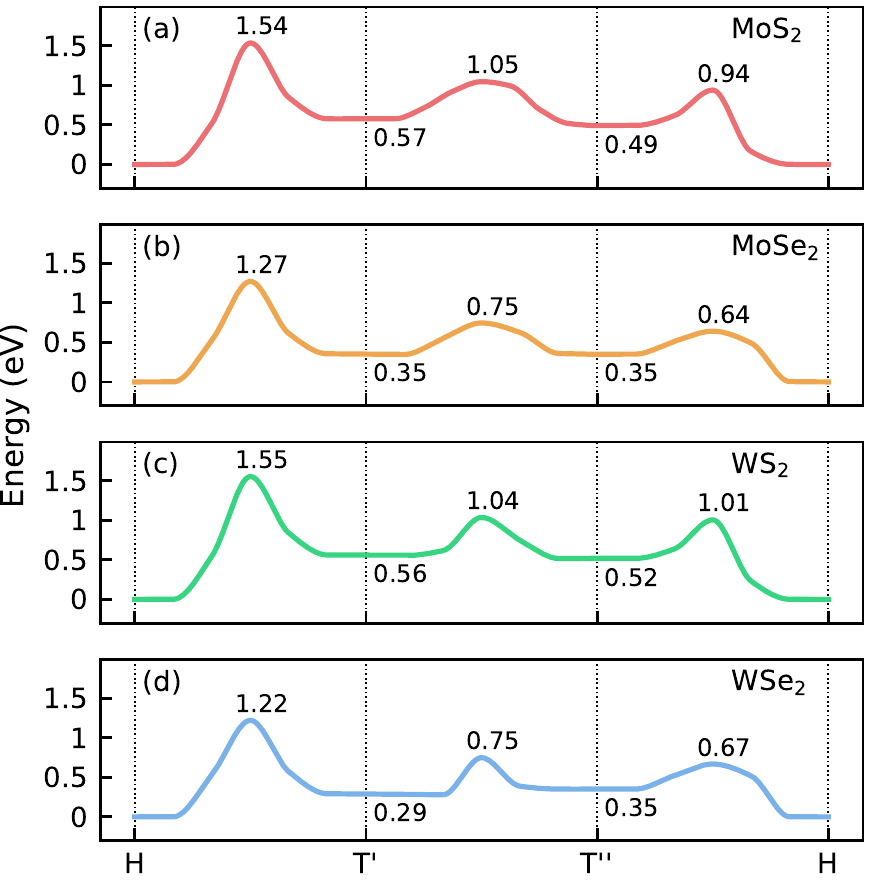}
\caption{\label{fig:neutral_en_barrier} The relative energy difference and transition barrier between the H phase, the \Tp\ phase and the \Tpp\ mixed phase of (a) \mos, (b) \mose, (c) \ws\ and (d) \wse. Energies are referred to that of the H phase for each material.}
\end{figure}

In agreement with previous DFT studies, we find the H phase to be the ground state structure for all four materials\cite{kretschmer2017structural,sun2016origin,wang2017structural,kwon2018intercalation,gao2015charge,nasr2015structural,kan2014structures}. The total energy differences between this phase and both the \Tp\ phase and the \Tpp\ mixed phase are given in Fig.~\ref{fig:neutral_en_barrier}. We find that the energy difference between the H phase and the \Tp\ phase is larger for the sulphides compared to the selenides. It reduces from 0.57~eV for \mos\ to 0.35~eV for \mose\ and from 0.56~eV for \ws\ to 0.29~eV for \wse. Similarly, the energy difference between the H phase and the \Tpp\ mixed phase reduces from 0.49~eV for \mos\  to 0.35~eV for \mose\ and from 0.52~eV for \ws\ to 0.35~eV for \wse. 
While the H phase is the ground state in all four cases, the \Tp\ phase is energetically more favourable than the \Tpp\ mixed phase for \wse\ but the ordering is opposite for the \mos\ and \wse\ (the difference is negligible for \mose).

The activation barriers to induce a phase transition from the H phase to the \Tp\ and \Tpp\ mixed phases, as calculated with the CI-NEB method, for all four materials are also shown in Fig.~\ref{fig:neutral_en_barrier}. 
In all cases, the energy barrier to transition directly from the H phase to the \Tp\ phase is large, ranging between 1.22~eV and 1.54~eV. In general, the barrier is smaller for \mse\ than for \ms. 
The energy barrier to transition from the H phase to the \Tpp\ mixed phase is smaller, ranging between 0.64~eV and 1.01~eV. Again, the smallest values occur for the \mse\ compounds. In all cases the energy barrier to transition from the ground state H phase to the lowest-lying T phase (namely, \Tpp\ for \mos, \mose\ and \ws, and \Tp\ for \wse) lies between 0.64~eV and 1.22~eV. Evidently, the H phase is very stable and will not convert to the T phase spontaneously, in agreement with experiment. Note that the transition from the \Tp\ phase to the \Tpp\ mixed phase requires the conversion of only half of the lattice from the \Tp\ phase to the H phase while keeping the other half fixed. Consequently, the barrier to transition from the \Tp\ phase to the \Tpp\ mixed phase is lower (by between 50 and 65\%) than the barrier to undergo the complete transition from \Tp\ phase to the H phase.

The calculated barriers reported here are in units of eV per formula unit (MX$_2$) to facilitate comparison between four materials and the total activation energy required for transition will scale with the number of formula units which transition simultaneously \cite{guo2015probing}. Note that the phase transition between the H phase and the T phase was shown to be diffusive, rather than a simultaneous phase transition of the entire monolayer. As a result, the activation barrier per f.u.~will be modified by the exact details of the nucleation process.

\subsection{Charged Monolayers}

Introducing extra charge to the TMD via an interaction with explicitly-modelled strong donor atoms bypasses issues related to static charging (discussed in Appendix). We consider two different concentrations of Li atoms adsorbed on the surface of the monolayers, namely Li$_{0.5}$MX$_2$ and Li$_{1}$MX$_2$. The stability of the alkali metal-adsorbed monolayer decreases at higher concentrations, and eventually becomes unstable for Li$_{1.5}$MX$_2$, in agreement with experiment.\cite{leng2016phase,li2017intermediate,papageorgopoulos1995li}

The adsorption sites were determined by calculating the total energy of lithium adsorbed on all of the unique candidate sites, shown as blue circles in Fig.~\ref{fig:structures}. These sites include those directly on top of the metal atom, on top of the chalcogen atoms and on the hollow site. In agreement with previous calculations in the literature \cite{hong2017direct}, we find that irrespective of the material or phase the lowest energy adsorption site is on top of the metal atom (The preferred adsorption position on the \Tp\ phase is shown with a dashed circle in Fig.~\ref{fig:structures}).
When increasing the Li concentration to LiMX$_2$, the second Li atom per unit cell adsorbs on the opposite surface, minimizing the electrostatic interaction between the two Li atoms. This configuration is 0.1~eV per formula unit lower in energy than that with both Li atoms adsorbed on the same side of the monolayer.
For the case of the \Tpp\ mixed phase, the lithium atoms are also positioned on top of the metal atoms. For Li$_{0.5}$MX$_2$, one lithium atom is placed in the H phase region and another in the \Tp\  phase region. For Li$_{1}$MX$_2$, the remaining lithium atoms were placed on top the metal atoms on the opposite side of the slab. 

The extra charge introduced by the Li atoms causes an expansion of the TMD lattice constants. These values are given in Table~\ref{tab:li_structures}. A Li concentration of Li$_{0.5}$MX$_2$ increases the lattice constants of the H phase by between 0.3 and 1.75\% compared to the neutral lattice. Similarly, Li adsorption increases the lattice constants of the \Tp\ phase by between 9\% and 3.23\%. Increasing the lithium concentration causes the lattice to expand further, and the expansion is higher in the selenides compared to the sulfides. Bader charge analysis \cite{tang2009grid} finds that, for both concentrations considered here, each Li atom donates approximately 0.82 electrons per formula unit. This value is independent of the material type and phase, indicating that the nature of the interaction is similar. 

\begin{table*}[ht!]\centering
	\ra{1.2}
	\setlength{\tabcolsep}{6pt}
	\begin{tabular}{@{\extracolsep{6pt}}lcccccccccccccc@{}}
		\hline \hline 
		\T \B   & \multicolumn{6}{c}{Li$_{0.5}$MX$_2$} & &  \multicolumn{6}{c}{LiMX$_2$}  \\
		\cline{2-7} \cline{9-14} 
		\T \B  & \multicolumn{2}{c}{H} &  \multicolumn{2}{c}{\Tp} & \multicolumn{2}{c}{\Tpp} & &  \multicolumn{2}{c}{H} &  \multicolumn{2}{c}{\Tp}  &  \multicolumn{2}{c}{\Tpp}   \\ \cline{2-7} \cline{9-14} 
		\T \B  \AA & a & b &  a & b & a & b & & a & b & a & b & a & b \\
		\cline{2-3} \cline{4-5} \cline{6-7} \cline{9-10} \cline{11-12} \cline{13-14} 
		\mos\ \T \B	&5.53 & 3.19&5.75 & 3.25&11.49&3.20& &5.68&3.23&5.9 &  3.29&11.79&3.22 \\
		\mose\ \T \B	&5.8 &  3.32&6. &   3.37&11.95&3.34& &5.98&3.36&6.12 & 3.46&12.2&3.38\\
		\ws\ \T \B	&5.51 & 3.18&5.74 & 3.26&11.47&3.19& &5.98 & 3.36&5.88 & 3.27&11.72&3.19\\
		\wse\ \T \B	&5.81 & 3.29&5.97 & 3.4 &11.90&3.33& &5.94 & 3.33&6.06 & 3.47&12.30&3.33\\
		\hline
	\end{tabular}
	\caption{\label{tab:li_structures}\,The lattice parameters a, b (in \AA) of the H, the \Tp, and the  \Tpp\ mixed phase of Li$_{0.5}$MX$_{2}$ and LiMX$_2$ in the orthorhombic unit cell.}
\end{table*}

\begin{figure}
\includegraphics[width=\columnwidth]{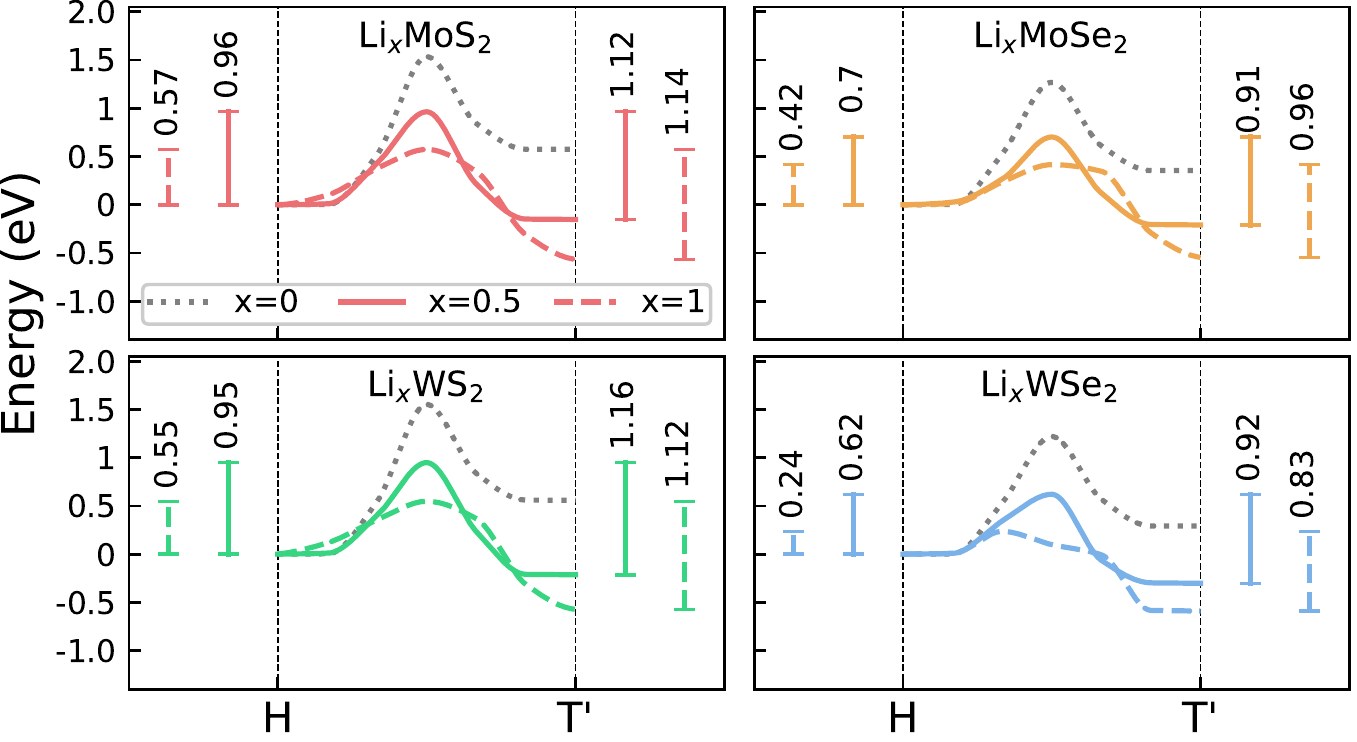}
\caption{\label{fig:charged_en_barrier}\,The relative energy difference and transition barrier between the H phase and the \Tp\ phase of Li$_x$\mos, Li$_x$\mose, Li$_x$\ws\ and Li$_x$\wse\ with $x$ = 0 (dotted line), $x$ = 0.5 (solid line) and $x$ = 1 (dashed line). Energies are referred to that of the H phase of each material.}
\end{figure}

The energy differences between the different polytypes, for all four compounds and for both Li concentrations, are summarized in Fig.~\ref{fig:charged_en_barrier}.
The \Tp\ phase becomes the ground state structure of all four Li-adsorbed TMDs. This is followed by the H phase and finally the \Tpp\ mixed phase. This is in agreement with previous studies on \mos\ which have shown a phase transition to occur at a Li concentration of  Li$_{0.4}$\mos\cite{enyashin2012density}, a K concentration of K$_{0.225}$\wse \cite{lei2018direct} and a Na concentration of Na$_{0.375}$\mos \cite{li2017intermediate}.

The energy difference between both the \Tp\ and \Tpp\ mixed phases compared to the H phase increases going down the group from S to Se and also going from Mo to W for both considered Li concentrations. On increasing the lithium concentration from Li$_{0.5}$MX$_2$ to Li$_{1}$MX$_2$, the energy difference between the \Tpp\ mixed phase and the H phase decreases slightly for the sulfides but increases for the selenides. In contrast to our observation that the energy of the \Tpp\ mixed phase increases with respect to the H phase, Ma et al. reported a threshold of 0.4 e per \mos\ to induce the transition from the H phase to the \Tpp\ mixed phase \cite{ma2016predicting}. This discrepancy could be a consequence of the static charging method used in obtaining those results.

These trends in stability can be explained by simple electron filling in the rigid band approximation \cite{lei2018direct}. For the H and \Tpp\ mixed phases, excess electrons cause an increase in total energy equal to the band gap. As the \Tp\ phase is semi-metallic, the next available energy level is at the Fermi level, and so this phase becomes lower in energy. 

The barriers for the H to \Tp\ phase transition in the charged monolayers are also shown in Fig.~\ref{fig:charged_en_barrier}. In all cases, the barrier for the phase transition decreases due to lithium adsorption.  
The maximum barrier for the transition from the H phase to the \Tp\ phase is found for \mos\ at a value of 0.96~eV for Li$_{0.5}$\mos\ and 0.57~eV for Li$_{0.5}$\mos. 
The minimum transition barrier from the H phase to the \Tp\ phase is found for \wse\ with a barrier of 0.62~eV for Li$_{0.5}$\wse\ and 0.24~eV for Li$_{1}$\wse.
\ws\ has the second highest barrier, at a value of 0.55~eV for the Li$_{1}$\ws\ structure, while \mose\ has a barrier of 0.42~eV at the same Li concentration. In all cases, the barrier to transition does not decease to zero. As such, the transition cannot be spontaneous and an energy equal to the barrier height needs to be provided to induce the transition. 

\section{Discussion}

Li adsorption is an exothermic process.
The adsorption energies, defined as $E_{\mathrm{ads}} = E_{\mathrm{Li}_x\mathrm{MX}_2} - E_{\mathrm{MX}_2} - x E_{\mathrm{Li}}$ where $E_{\mathrm{Li}_x\mathrm{MX}_2}$ is the total energy of the Li adsorbed structure, $E_{\mathrm{Li}_x\mathrm{MX}_2}$ is the total energy of the pristine monolayer, $E_{\mathrm{Li}}$ is the energy of an isolated Li atom and $x$ refers to the Li concentration, for all four materials are given in Table~\ref{tab:ads}. The Li adsorption energy is higher for \Tp\ phase as compared to the H phase. 
For Li$_{0.5}$MX$_2$, the adsorption energy is largest for \wse\ at $-1.03$~eV followed by \mose\ and \mos\ at $-0.99$ and $-0.97$~eV per formula unit respectively, and smallest  for \ws\ at $-0.85$~eV per formula unit.  This energy is comparable to the energy barrier to the phase transition at this Li concentration. For Li$_{1}$MX$_2$, the adsorption energy per MX$_2$ approximately doubles, to at least 2.5 times the energy barrier. If some of this energy is used to overcome the barrier to the transition, Li adsorption may be sufficient to make the process spontaneous. 
Furthermore, given that the rate of transition decreases exponentially with activation energy (as per the Arrhenius equation), we can conclude that, for a given X, \wx\ will transition at a slightly higher rate than \mox. Likewise, for a given M, \mse\  will transition at a higher rate than \ms.

\begin{table}\centering
	\ra{1.2}
	\setlength{\tabcolsep}{15pt} % Default value: 6pt
	\begin{tabular}{@{\extracolsep{3pt}}lcccc@{}}
		\hline \hline
		\T \B   & \multicolumn{2}{c}{Li$_{0.5}$MX$_2$} &   \multicolumn{2}{c}{LiMX$_2$}  \\
		\cline{2-3} \cline{4-5} 
		\T \B  & E$_{ads}^{\mathrm{H}}$ & E$_{ads}^{\mathrm{T}^\prime}$ & E$_{ads}^{\mathrm{H}}$ & E$_{ads}^{\mathrm{T}^\prime}$   \\
		\cline{2-3} \cline{4-5} 
		\mos\ \T \B &	-0.97	&	-1.64	&	-2.51	&	-3.24	\\
		\mose\ \T \B &	-0.99	&	-1.41	&	-2.36	&	-2.89	\\
		\ws\ \T \B &	-0.85	&	-1.58	&	-2.26	&	-2.98	\\
		\wse\ \T \B &	-1.03	&	-1.38	&	-2.16	&	-2.70	\\
		\cline{1-5}
	\end{tabular}
	\caption{\label{tab:ads}\, Adsorption energy, $E_{\mathrm{ads}}$, of lithium on the H and \Tp\ TMD phases (eV per unit MX$_2$).}
\end{table}

This is contradictory to the observation of a higher percentage of \Tp\ phase found for \ws\ compared to \mose\ and \wse. To explain this, we must also consider the fact that in order to measure the T/H ratio experimentally, the exfoliated monolayers are washed with deionized water and dried in vacuum. In this process the Li ions desorb and the monolayers revert to their neutral state. They are prevented from transitioning immediately back to the ground-state H phase by an energy barrier. This is supported by the observation of the M$^{4+}$ oxidation state in X-ray photoelectron spectra (XPS) \cite{li2017intermediate,ambrosi20152h,rohaizad20171t}. 

The barriers for the reverse transition from the neutral \Tp\ phase to the neutral H phase, shown in Fig.~\ref{fig:neutral_en_barrier}, are now relevant. For all materials, the barrier to return directly to the H phase is prohibitively high, ranging between 0.90~eV and 0.99~eV. Instead, the barrier to transition to a mixed \Tpp\ type phase is significantly lower, ranging between 0.39~eV and 0.48~eV. Experimental evidence for such an indirect transition from the \Tp\ phase back to the H phase for \mos\ via a mixed \Tpp\ mixed phase can be found: By fitting the rate equation to in-situ Raman measurements, the barrier to transform from the meta-stable metallic phase to the ground state was found to be 400~meV \cite{guo2015probing}. This barrier value is consistent with the calculated barrier to transition from the \Tpp\ mixed phase to the H phase of 450~meV calculated here (cf.~Fig.~\ref{fig:neutral_en_barrier}) whereas the barrier to transition from the \Tp\ phase back to the H phase directly is 970~meV. This suggests that the transition from the \Tp\ phase to the H phase occurs in two steps for this material, via the \Tpp\ mixed phase.

The barrier to transition from the \Tpp\ mixed phase to the H phase is smallest for the two selenium compounds at 0.23~eV and 0.27~eV for \mose\ and \wse, respectively. The barrier for \ws\ is twice as high, at 0.49~eV. The significance of this difference in barrier for \ws\ and \wse\ can be determined by calculating the exponential factor of the Arrhenius equation at room temperature ($k_{B}T = 25.7 \mathrm{~meV}$). This is of the order of ~$5\times10^{-9}$ for \ws, whereas it is of the order of  ~$1\times10^{-4}$ for \wse. This rate difference is significant and its effects should be experimentally observable. This can explain the significant difference between the percentage of \Tp\ phase observed in \ws\ compared to both \mose\ and \wse,~\cite{ambrosi20152h}. However, the calculated activation barriers cannot explain the low 1T/2H ratio found in exfoliated \mos. This discrepancy could be due to non-electronic effects, such as a poor exfoliation efficiency in certain organometallic compounds, edge effects related to the change in lateral size of the exfoliated monolayers, the presence of defects or the oxidation of the exfoliated layers. Another possibility is that the simplified phase transition mechanism considered here does not consider the full nucleation process. 

Finally, we note that the \Tpp\ mixed phase comprises of alternating H and \Tp\ phases, each a unit cell thick, as illustrated in Fig.~\ref{fig:structures}. However, this phase can be viewed as just one example of a family of mixed \Tp\ and H phase structures. By changing the \Tp:H ratio in \mos\ from 1:1 for the \Tpp\ structure to 2:1, the total energy difference between it and the H phase increases from 0.49~eV to 0.51~eV. By increasing the ratio further to 3:1, the energy difference increases further to 0.52~eV. Clearly, the total energy of \mos\ is strongly dependent upon the fraction of H phase. The energy of all such families of confined structures will be higher than that of the H phase but lower than that of the \Tp\ phase. The exact composition of such confined structures will be dependent on the available energy and the TMD flake size and shape and are not confined to the idealised one dimensional structures discussed here. Experimentally, the boundary between the H phase and \Tp\ phase areas are observed to be atomically sharp, as in the \Tpp\ mixed phase. This boundary evolves over time via a transversal displacement of one of the S planes~\cite{lin2014atomic,diffuse2018} leading to the complete phase transition of the flake. The observed partial phases, with mixed metallic (\Tp) and insulating (H) regions, can be understood as intermediate stable structures which have lower barriers of transition.

\section{Conclusion}

Group-VI TMDs are known to undergo structural phase transitions from a semi-conducting H phase to a metallic T phase when subjected to alkali-metal assisted exfoliation. The efficiency of this process is strongly dependent on the chemical composition of the material.
We show that the ratio of \Tp\ to H phase is maximized if the charge-induced transition from the H phase to the \Tp\ phase is favourable and the reverse transition upon removal of charge is unfavorable. For example, the high proportion of T phase found in WS$_2$ monolayers after alkali treatment can be explained by a high barrier to revert back to the H phase after the initial phase transition has been induced. 
While charged \mse\ materials have the lowest energy barriers to the phase transition, the barriers to revert back to the H phase are also low. This can explain the relatively low content of metallic phase found in \mose\ and \wse\ after exfoliation.
Finally, the low proportion of metallic phase found in \mos\ monolayers in some experiments cannot be explained using the activation barriers alone and so non-electronic effects, such as a differing exfoliation efficiency or differing levels of monolayer oxidation, must also contribute to the outcome.\\

\begin{acknowledgments} 
This work was supported by a Science Foundation Ireland Starting Investigator Research Grant (15/SIRG/3314). 
Computational resources were provided by the supercomputer facilities at the Trinity Center for High Performance Computing (TCHPC) and at the Irish Center for High-End Computing (project tcphy091b). 
\end{acknowledgments}

\appendix
\section{Implicit Charging} \label{static}
We highlight here some issues related to statically charging monolayers or slabs in DFT with periodic boundary conditions, as it is to these issues that we attribute the wide range of values reported in the literature for the critical charge required to induce a structural phase transition in TMDs.

For the case of isolated charged slabs, such as the TMD monolayers considered here, the electric field due to the extra uniform charge density is constant and should result in potential that varies linearly with distance from the surface of the slab. When periodic boundary conditions are implemented, this linear potential is replaced by the combined effective potential due to the consecutive periodic images. The quantities relating to the isolated slabs can be recovered from these periodic calculations by applying a correction term to the effective potential and total energy. The exact functional form of this potential is discussed in detail by  Andreussi at al.~\cite{andreussi2014electrostatics} As discussed in manual for {\sc vasp} (in the section Monopole, Dipole and Quadrupole corrections) the leading term of this correction, which cancels the interaction of the linear potential with the background compensating charge, is absent. As a result, the total energy is essentially incorrect and cannot be relied upon.

Furthermore, we find that above a certain critical charge, positive eigenvalues are occupied. This was previously shown to occur for atomic anions and can be attributed to the self-interaction error~\cite{kim2011communication}.

\begin{figure}[!h]
	\centering
	\includegraphics[width=\columnwidth]{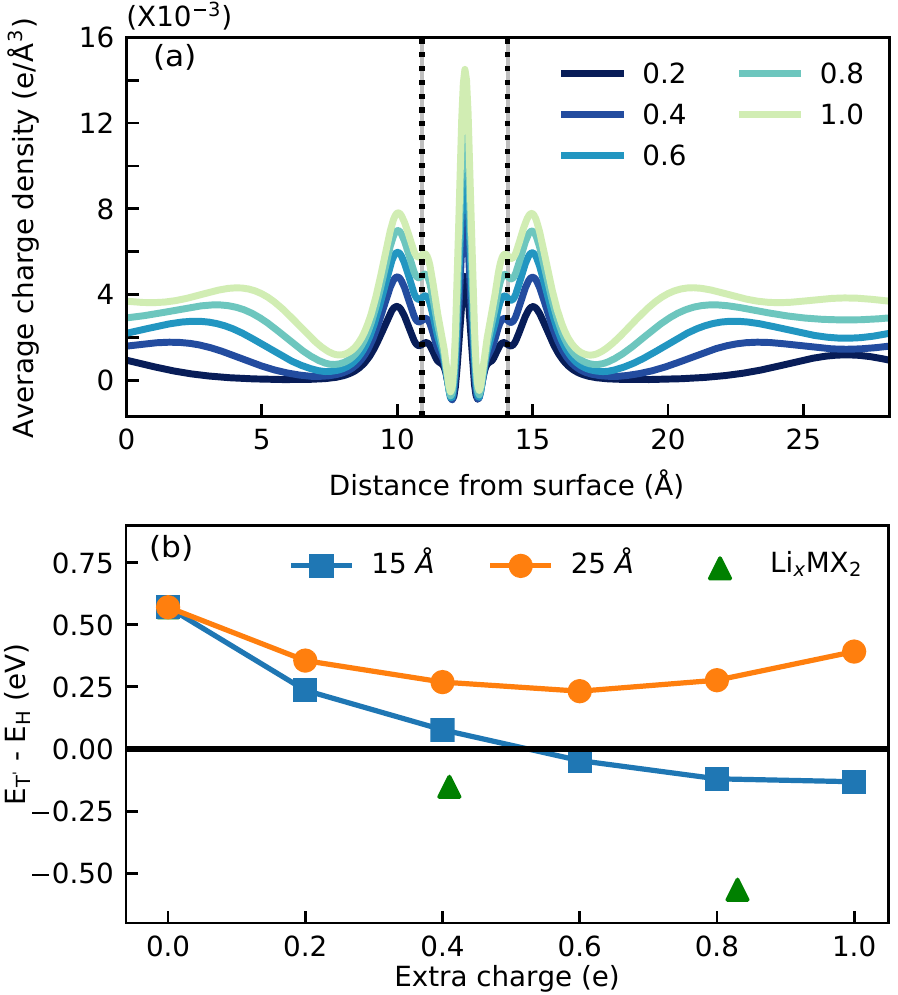}
	\caption{\label{fig:static} (a) Planar average of the charge density across the TMD monolayers showing charge leakage into the vacuum region for typical values of excess charge. The positions of the chalcogen atoms are marked by black dashed lines. (b) The energy difference between the \Tp\ and the H phase of \mos\ as a function of excess static charge for two different vacuum thickness, compared to the values found by explicitly charging the slab using Li atoms.}
\end{figure}

Finally, Topsakal et al.~showed that, when the basis set is present in the vacuum region, excess negative charge does not stay on the slab but spills over in the vacuum, beyond a critical value which depends on the vacuum thickness~\cite{topsakal2012effects}. This phenomenon is demonstrated in Fig.~\ref{fig:static}(a)  for the H phase of \mos\ with a vacuum length of 25~\AA\ perpendicular to the surface of the slab and was previously shown for graphene~\cite{topsakal2011static}. The excess charge density added is placed in the vacuum and on the outer faces of the monolayer rather than on the \textit{d} orbital of the metal atom.

To illustrate how these issues affect the determination of the critical charge required to induce a structural phase transition, we show in Fig.~\ref{fig:static}(b) the total energy difference between the H and the \Tp\ phase of \mos\ as a function of excess charge, and for two different vacuum lengths. All three sources of error are now present, namely charge has spilled into the vacuum (above a certain critical excess charge value), the appropriate monopole correction is absent and positive eigenvalues are occupied. We find that the vacuum length of 15~\AA\ (blue squares) shows a phase transition for excess charge of around 0.5 e per f.u. On increasing the vacuum length to 25~\AA\ (orange circles) no such transition is observed. 

Clearly, it is not possible to get physically meaningful results using statically charged slabs combined with periodic boundary conditions in density functional theory. A similar erroneous dependence on the vacuum length was also reported by Bal et al.~for adsorbed molecules on charged surfaces, further highlighting the irreproducibility of such results\cite{bal2018modelling}.

\FloatBarrier

\end{document}